\documentclass{article}
\usepackage[utf8]{inputenc}
\usepackage{authblk}

\providecommand{\PACS}[1]{\textbf{PACS} #1}

\providecommand{\acknowledgments}[1]{\textbf{Acknowledgments:} #1}

\usepackage{amsmath}
\usepackage{graphicx}
\graphicspath{ {./images/} }
\usepackage{hyphenat}
\usepackage{latexsym}
\usepackage{stackrel,amssymb}
\usepackage{esint}

\title{Inflationary solutions in the simplest gravity model with conformal symmetry}
\author[1,2]{S. Alexeyev }
\author[1,3]{D.Krichevskiy}
\affil[1]{Sternberg Astronomical Institute, Lomonosov Moscow State University}
\affil[2]{Department of Quantum Theory and High Energy Physics, Faculty of Physics, Lomonosov Moscow State University}
\affil[3]{Department of Physics, Bauman Moscow State Technical University}
\date{}

\begin{document}

\maketitle

\begin{abstract}
    We discuss a model of gravity with conformal symmetry appearing in the simplest extension of General Relativity with the Poincaré algebra terms. The nonlinear realization of symmetry causes the existence of five scalar fields. Therefore it looks desirable to use them for driving the inflation at the earliest stages of the Universe evolution. It is shown that the evolution of the scale factor doesn't imply accelerated expansion so that there are no inflationary solutions in this model. To drive inflation a more complicated model induced by extension of the Poincare algebra is required.
\end{abstract}


\section{Introduction}
During last years the problems of dark matter and dark energy attracted a lot of attention. One of the ways to model these phenomena is to extend the General Relativity (GR) by additional fields or corrections. Among the others the idea to use the conformal symmetry looks very perspective. Following this approach the study of the nonlinear realization of conformal symmetry was carried out  \cite{Pervushin1,Pervushin2,Pervushin3}. In this context the new model of gravity where General Relativity (GR) was extended by nonlinear conformal symmetry has been suggested in  \cite{Arbuzov2020}. Indeed, the nonlinear symmetry realization is the other name of spontaneous symmetry breaking (except the special case with Abelian group which is not nonlinear). Here it is necessary to note that a symmetry group $G$ of a physical system is spontaneously broken down to a subgroup $H$ if the system state is invariant under group $H$ but not invariant with respect to $G$ \cite{RubakovGaugeFields}. Higgs sector from the standard model can serve as an example: its potential is invariant under $SU(2) \times SU(2)$ group, but any ground state is not. So the existence of the bigger symmetry group $G$ is a fundamental property of a system even if $G$ is broken down to $H$. 

Manifestations of the bigger symmetry are nonlinear to include the partial symmetry from the subgroup $H$. The notion of spontaneous symmetry breaking is used to emphasize that the matter respects only a local symmetry group, while the concept of nonlinear symmetry realization is used to note that the original bigger symmetry group still manifests itself. The fact that nonlinear symmetry realizations and spontaneous symmetry breaking refer to the same phenomenon allows to apply the Goldstone theorem. Therefore if a global symmetry group $G$ is realized on a subgroup $H$ and a quotient $G/H$ the correspondent model contains $dim(G)-dim(H)$ massless degrees of freedom, i.e. Goldstone bosons \cite{Nair2005}. As a result massless gravitons could be interpreted as Goldstone bosons hence the gravity itself appears due to a nonlinear manifestation of some symmetry.

A consistent quantum treatment of gravity in the low-energy regime becomes possible within the effective field theory (EFT) formalism. EFT describes all observed phenomena in terms of low-energy effective gravitons' interactions. Therefore treating of GR as EFT provides a ground to conjecture that gravity behavior in low energy regime is managed by a spontaneous breaking of some symmetry. Therefore, in low-energy regime effective massless (in accordance with the empirical data \cite{Abbott161101}) gravitons can be treated as Goldstone bosons. GR contains only one mass parameter, namely the Planck mass which should be considered as the symmetry-breaking scale. 
For the implementation of the nonlinear symmetry realization to gravity as a first step one has to define groups $G$ and $H$. Gravitons are described within GR by the metric tensor $g_{\mu\nu}$ with ten independent components. To provide the same number of components it is necessary to use ten-dimensional Poincare group $P$ as the subgroup $H$. 

A proper way to define the group $G$ is suggested by the Ogievetsky theorem: {\it all the generators of coordinate frame transformations can be obtained by the commutation of conformal group $C(1, 3)$  and special linear one $SL(4, \mathbb{R})$ generators} \cite{Ogievetsky1979}. Thus, in \cite{Arbuzov2020} $C(1, 3)$ or $SL(4, \mathbb{R})$ were considered as perspective candidates for $G$. Further algebras of $C(1, 3)$ and $SL(4, R)$ extend the Poincare one with three operator sets: $R_{(\mu)(\nu)}$, $K_{(\mu)}$ and $D$. We study a model induced by the simplest expansion of the Poincare algebra with operators $D$ and $K_{(\mu)}$ generating five new physical fields: $\psi$ and $\sigma^{(\alpha)}$, $\alpha=\overline{1,4}$ with the canonical mass dimension. Note that the index $(\alpha)$ is not the Lorentz one so $\sigma^{(\alpha)}$ represents four scalar fields connected with nonlinear transformations.

The discussed model attaches some interest because of the problem of the inflation origin. A scalar field (inflaton) is responsible for the inflation in the early Universe. Excitations of this inflaton field are expected to be quantified and the corresponding particles must be zero-spin ones. However, in GR there are no any spin-zero elementary particles except the Higgs boson. Therefore models containing additional scalar fields become  interesting. In \cite{Arbuzov2020} it was supposed that $\psi$ and $\sigma^{(\alpha)}$ can serve as inflaton fields.

\section{Inflation in the model}
\label{sct:2}

The Lagrangian with additional scalar fields $\sigma^{(\alpha)}$ and $\psi$ is suggested in \cite{Arbuzov2020}:
\begin{equation} \label{eq:1}
    \mathcal{L} = \frac{1}{2} f_1\left( \partial\psi  \right)^2 +\frac{1}{2} f_2 \left( \partial \sigma  \right)^2 + f_3 \partial_{\mu}\psi \partial^{\mu}\sigma^{\left( \alpha \right)}\sigma_{\left( \alpha \right)},
\end{equation}
    where 
\begin{eqnarray*}
    f_1 & = & 1 +\frac {\sigma^2}{\varepsilon^2} \left[ \frac{e^{\psi /\varepsilon} - \psi /\varepsilon-1 }{ \left( \psi /\varepsilon  \right)^2}  \right]^2 , \\ 
    f_2 & = & \left( \frac { e^{\psi /\varepsilon}-1 }{ \psi /\varepsilon  }  \right)^2, \\
    f_3 & = &  -\frac{1}{\varepsilon} \frac{e^{\psi /\varepsilon  }-1 }{ \psi /\varepsilon  } \frac { e^{ \psi /\varepsilon  }-\psi /\varepsilon -1 }{ \left( \psi /\varepsilon  \right)^2 }, \\ 
    \sigma^2 & = & \sigma^{\left(\alpha\right)} \sigma_{\left(\alpha\right)}, \\ 
     \left(\partial\psi\right)^2 & = & \partial_\mu\psi\partial^\mu\psi, \\ \left(\partial\sigma\right)^2 & = &  \partial_\mu\sigma^{\left(\alpha\right)}\partial^\mu\sigma_{\left(\alpha\right)}, 
\end{eqnarray*}
$\mu,\nu=\overline{0,3}$ and $\varepsilon$ is characteristic energy scale of the theory (i.e. symmetry breaking scale). It is important that the fields $\psi$, $\sigma^{(\alpha)}$ may serve as the origin of the inflation in the early Universe. Inflation stage is characterized by the accelerated expansion so that $\dot{a}>0$, $\ddot{a}>0$ \cite{Weinberg}, where $a(t)$ is the scale factor and dot denotes derivative over $t$.

On the inflation stage the Universe can be described by the following metrics \cite{Weinberg} 
\begin{equation} \label{eq:2}
    {ds}^2={dt}^2-a^2(t)({dx}^2+{dy}^2+{dz}^2),
\end{equation}
therefore, Ricci tensor $R_{\mu\nu}$ is diagonal:
$$
 R_{ \mu  \nu }=diag \left( -3\frac{\ddot{a}}{a}\text{, 2}\dot{a}^{2}+a\ddot{a}\text{, 2}\dot{a}^{2}+a\ddot{a}\text{, 2}\dot{a}^{2}+a\ddot{a} \right).
$$
The symmetric stress-energy tensor for the model (\ref{eq:1}) can be easily calculated and has the form:
$$
    \begin{gathered}
        T_{ \mu \nu  }=\frac { \partial { \mathcal{L} } }{ \partial \left( \partial^{ \mu  }\psi \right)  } \partial_{ \nu  }\psi +\frac { \partial { \mathcal{L} } }{ \partial \left( \partial^{ \mu  }\sigma^{ (\alpha ) } \right)  } \partial_{ \nu  }\sigma^{ (\alpha ) }-g_{ \mu \nu  }{ \mathcal{L} }=\\{ f }_{ 1 }\partial_{ \nu  }\psi \partial_{ \mu  }\psi +{ f }_{ 2 }\partial_{ \mu  }\sigma_{ (\alpha ) }\partial_{ \nu  }\sigma^{ (\alpha ) }+2{ f }_{ 3 }\partial_{ (\nu  }\psi \partial_{ \mu ) }\sigma_{ (\alpha ) }\sigma^{ (\alpha ) }-{ g }_{ \mu \nu  }{ \mathcal{L} }
    \end{gathered}
$$
where Lorenz index brackets denote symmetrization procedure. Evolution of scale factor is determined by Einstein field equations \cite{Alexeyev}
$$
   R_{ \mu  \nu }=k \left( T_{ \mu  \nu }-\frac{1}{2}T g_{ \mu  \nu } \right) , 
$$
where $T=T_{ \nu }^{ \nu }$ is the trace of stress–energy tensor,  $k=8 \pi G$ and $G$ is Newtonian gravitational constant. In the homogeneous case $\partial_i \psi =\partial_i \sigma_{(\alpha )} = 0$, $i=\overline{1,3}$, ten Einstein equations reduce to a system of only two following equations 
$$
\begin{gathered}
-3\frac { \ddot { a }  }{ a }=5k(4{ f }_{ 1 }{ \dot { \psi  }  }^{ 2 }+{ f }_{ 2 }\dot { \sigma  }^{ 2 }+{ f }_{ 3 }\dot { \sigma  }_{ (\alpha ) }\sigma^{ (\alpha ) })\\ 2\dot { a }^{ 2 }+a\ddot { a } =0,
\end{gathered}
$$
from which follows that 
\begin{equation} \label{eq:3}
   a(t)=c_{2}(c_{1}+3t)^{\frac{1}{3}},  
\end{equation}
and 
\begin{equation}\label{eq:3}
    \ddot{a} =-2\frac{\dot{a}^{2} }{a} <0
\end{equation}
where $c_{1}$ and $c_{2}$ are integration integration constants.
If the inflation stage is realized than $\ddot{a}(t)$ should be positive for any $t\in [t_i,t_f]$, where $t_f$ is the inflation end time, in addition for the expanding Universe $\dot{a}(t) > 0$, which contradicts with (\ref{eq:3}). Thus the model (\ref{eq:1}) doesn't provide stress energy tensor which leads to accelerated expansion.

\section{Conclusions}
\label{sct:3}

The general conclusion of our analysis is that the inflation stage doesn't exist in the model (\ref{eq:1}) at least if spacial perturbations are not considered. Such situation becomes real due to the fact that the Lagrangian (\ref{eq:1}) emerged by spontaneous symmetry breaking contains only the derivatives of scalar fields. 

It is important to point out that this result looks very significant in the context of conformal gravity models. Note that the theory with nonlinear realization of a conformal symmetry (\ref{eq:1}) appearing to become a new very perspective class among conformal gravity ones. The rigorous proof that inflation is absent in this model means that, unfortunately, it can't be used in realistic cosmological scenarios. Hence the only model where Poincare algebra is extended by operators $R_{(\mu)(\nu)}$ has chances to be compatible with reality and our conclusion significantly constraints the set of such theories. Further the discussed model corresponds to the leading perturbative order, i.e. to the tree-level one. Note that the well-known  Starobinsky   inflation \cite{Starobinsky} occurs due to radiation corrections. Thus, one can't rule out that in the discussed model scalar fields won't get any effective potential $V_{eff}$ as in \cite{Arbuzov2020_1} there are indications to the possibility of such a scenario implementation. That is why we suppose that the fact that the inflation is absent at the tree level is important by itself. Of course, a detailed study of a model induced by the extension of the Poincare algebra with operators $R_{(\mu)(\nu)}$ is required as the next step in conformal gravity development.

\vspace{6pt}

\acknowledgments{The authors would like to thank Dr. Boris Latosh for the useful discussions and critical notes on the subject of the paper. They also thank a scientific educational school of MSU “Fundamental and applied space research”.}


\begin{thebibliography}{999}

\bibitem{Pervushin1}
Arbuzov A.B., Barbashov B.M., Borowiec A., Pervushin V.N., Shuvalov S.A., Zakharov A.F. {General relativity and the standard model in scale-invariant variables}{\em Grav. Cosmol.}, {\bf 2009}. V.15, P.199-212. 

\bibitem{Pervushin2}
Arbuzov A.B., Barbashov B.M., Nazmitdinov R.G., Pervushin V.N., Borowiec A., Pichugin K.N., Zakharov A.F. {Conformal Hamiltonian Dynamics of General Relativity} {\em Phys. Lett. B}, {\bf 2010}. V.691, no.5, P.210-233. 

\bibitem{Pervushin3}
Pervushin V.N., Arbuzov A.B., Barbashov B.M., Nazmitdinov R.G., Borowiec A., Pichugin K.N., Zakharov A.F. {Conformal and Affine Hamiltonian Dynamics of General Relativity.}
  {\em Gen. Rel. Grav.}, {\bf 2012}. V.44, P. 2745–2783. 


\bibitem{Arbuzov2020}
Arbuzov A., Latosh B. {Gravity and nonlinear symmetry realization} {\em Universe}, {\bf 2020}.  V.6, no.1, P. 12. arXiv:1904.06516


\bibitem{RubakovGaugeFields}
Rubakov V.A. {Classical Theory of Gauge Fields}.{\em Princeton University Press}, {\bf 2009}.

\bibitem{Nair2005}
Nair V.P. {Quantum Field Theory: A Modern Perspective; Graduate Texts in Contemporary Physics}.{\em Springer:New York}, {\bf 2005}.

\bibitem{Abbott161101}
Abbott B.P.; et al. [LIGO Scientific and Virgo Collaboration]  {GW170817: Observation of Gravitational Waves from a Binary Neutron Star Inspiral.} {\em Phys. Rev.D.}, {\bf 2017}. V.119, no.16,  P.161101. 

\bibitem{Ogievetsky1979}
Ogievetsky V.I. {Infinite-dimensional algebra of general covariance group as the closure of finite-dimensional algebras of conformal and linear groups}{\em Lett. Nuovo Cimento}, {\bf 1973}. V.8, no.17, P.~988–990. 

\bibitem{Weinberg}
Weinberg S. {Cosmology}. {\em Oxford University Press}, {\bf 2008}.

\bibitem{Alexeyev}
Alexeyev S.O., Pamyatnykh E.A., Ursulov A. V., Tretyakova D.A., Latosh, B.N. {General relativity: Introduction. Modern development and applications}. {\em URSS}, {\bf 2020} (in Russian).

\bibitem{Starobinsky}
Starobinsky A.A. A new type of isotropic cosmological models without singularity {\em Phys. Lett. B},
{\bf 1980}. V.91, no.1, P.99–102. 


\bibitem{Arbuzov2020_1}
Arbuzov A., Latosh B. Effective potential of scalar-tensor gravity. arXiv:2007.06306.


\end{thebibliography}
\end{document}